\documentclass[pra,amsfonts,showpacs,preprint,showkeys]{revtex4}
\usepackage[T1]{fontenc}
\usepackage{graphicx}
\RequirePackage{times}
\RequirePackage{mathptm}
\begin{document}
\title{Quantum Theory Looks at Time Travel}

\author{Daniel M. Greenberger}
\affiliation{Department of Physics, City College of the City University of New York     \\
  New York, NY 10031, USA}
\email{greenbgr@sci.ccny.cuny.edu}
\author{Karl Svozil}
\affiliation{Institute of Theoretical Physics, Vienna
    University of Technology, Wiedner Hauptstra\ss e 8-10/136, A-1040
    Vienna, Austria}
\email{svozil@tuwien.ac.at}

\begin{abstract}
We introduce a quantum mechanical model of time travel which includes two figurative beam splitters in order to induce feedback to earlier times. This leads to a unique solution to the paradox where one could kill one's grandfather in that once the future has unfolded, it cannot change the past, and so the past becomes deterministic. On the other hand, looking forwards towards the future is completely probabilistic. This resolves the classical paradox in a philosophically satisfying manner.
\end{abstract}


\maketitle

Classically, time travel is inconsistent with free will. If one could
visit the past, then one could change the past, and this would lead
to an alternative present. So there is a paradox here, which is best
illustrated by the famous scenario of a person going back in time
to shoot his father before his father has met his mother, and thus
negating the possibility of his having ever been born. It is for reasons
like this that time travel has been considered impossible in principle
{[}1{]}.

Of course, one can get around this problem if one considers the universe
to be totally deterministic, and free will to be merely an illusion.
Then the possibility of changing the past (or the future, for that
matter) no longer exists. Since we prefer to think that the writing
of this paper was not preordained at the time of the big bang, we
shall reject this solution on psychological grounds, if not logical
ones, and ask whether the paradoxes of classical physics can be gotten
around, quantum mechanically.

Most attempts to go beyond the confines of classical theory in order
to study time travel have been in the framework of relativity theory,
making use of freedom to warp the topological properties of spacetime.
We shall not comment on these here, except to note that they are not
incompatible with what we shall be saying, and might conceivably be
combined with it.

It seems to us that time travel is very much in the spirit of quantum
mechanics, and in fact, it seems quite arbitrary and outside the spirit
of the subject to forbid it {[}2{]}. For example, if one studies the
propagation of a physical system from time $t_{1}$ to later time
$t_{2}$, one writes
\begin{equation}
\psi\left(t_{2}\right)=U\left(t_{2},t_{1}\right)
\psi\left(t_{1}\right),\,\,\,\,\,\,\, t_{2}>t_{1},\,\,\,\label{4.1}\end{equation}
where $U$ is some unitary operator describing the dynamical unfolding
of the system. To calculate $U$, some sums over all possible paths
leading from the initial state to the final state, but restricting
these paths to the forward direction of time.

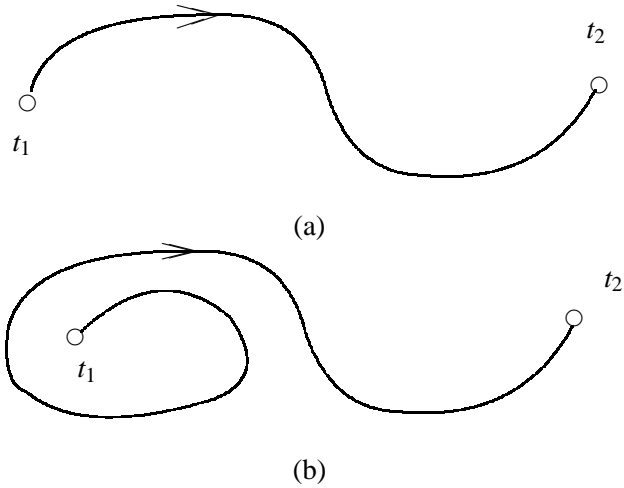
\begin{figure}
\begin{center}
\unitlength 1.00mm
\linethickness{0.7pt}
\begin{picture}(79.00,23.00)
\bezier{132}(2.33,11.67)(5.33,22.00)(27.33,21.67)
\bezier{92}(27.33,21.67)(39.00,22.00)(41.66,11.34)
\bezier{80}(41.66,11.34)(45.00,0.67)(54.00,0.34)
\bezier{124}(54.00,0.34)(70.33,-1.33)(77.33,11.67)
\put(2.00,10.00){\circle{2.00}}
\put(78.00,12.33){\circle{2.00}}
\put(1.33,4.33){\makebox(0,0)[cc]{$t_1$}}
\put(77.67,19.33){\makebox(0,0)[cc]{$t_2$}}
\put(27.35,21.66){\line(-4,1){5.39}}
\put(27.35,21.66){\line(-3,-1){5.39}}
\end{picture}
\\
(a)
\\

\unitlength 1.00mm
\linethickness{0.7pt}
\begin{picture}(81.00,23.75)
\bezier{132}(0.33,12.67)(3.33,23.00)(25.33,22.67)
\bezier{92}(25.33,22.67)(37.00,23.00)(39.67,12.33)
\bezier{80}(39.67,12.33)(43.00,1.67)(52.00,1.33)
\bezier{124}(52.00,1.33)(68.33,-0.33)(75.33,12.67)
\bezier{44}(0.33,12.67)(-0.67,5.00)(2.67,4.00)
\bezier{112}(2.67,4.00)(10.33,-2.33)(27.67,3.00)
\bezier{72}(27.67,3.00)(35.33,6.00)(29.67,14.00)
\bezier{104}(29.67,14.00)(20.33,21.67)(10.00,12.00)
\put(9.33,11.23){\circle{2.00}}
\put(75.63,13.73){\circle{2.00}}
\put(11.00,6.67){\makebox(0,0)[cc]{$t_1$}}
\put(81.00,18.67){\makebox(0,0)[cc]{$t_2$}}
\put(25.35,22.65){\line(-4,1){4.39}}
\put(25.35,22.65){\line(-4,-1){4.39}}
\end{picture}
\\
(b)
\end{center}
\caption{
In the path integral one can take all paths (a) that go
forward in time, but one excludes all paths (b) that go backward in
time.
\label{f4.1}}
\end{figure}

Furthermore, it is well known that when one makes measurements in
quantum theory, one's simple sense of causality is violated, and so
a classical sense of causality is a rather poor guide as to what should
or should not be allowed quantum mechanically. And this restriction
would seem to violate the spirit of the entire enterprise. Specifically,
why should there not be some form of feedback into the past in determining
what will happen in the future (see Fig.~\ref{f4.1})?

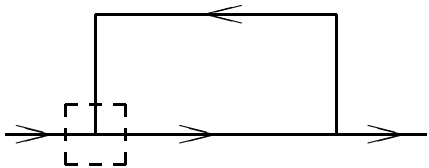
\begin{figure}
\begin{center}

\unitlength 0.80mm
\linethickness{0.7pt}
\begin{picture}(65.00,26.49)
\put(0.00,5.00){\line(1,0){70.00}}
\put(15.00,5.00){\line(0,1){20.00}}
\put(15.00,25.00){\line(1,0){40.00}}
\put(55.00,25.00){\line(0,-1){20.00}}
\put(10.00,0.00){\dashbox{2.00}(10.00,10.00)[cc]{}}
\put(35.03,4.99){\line(-4,1){5.99}}
\put(35.03,4.99){\line(-4,-1){5.99}}
\put(66.36,4.99){\line(-4,1){5.99}}
\put(66.36,4.99){\line(-4,-1){5.99}}
\put(7.86,4.99){\line(-4,1){5.99}}
\put(7.86,4.99){\line(-4,-1){5.99}}
\put(33.30,24.99){\line(4,1){5.99}}
\put(33.30,24.99){\line(4,-1){5.99}}
\end{picture}
\end{center}
\caption{
In classical feedback circuit, one inserts a loop that goes
from a later time to an earlier time. The loop then has two entry
ports and only one exit port, so that one cannot uniquely reverse
it, and if tried quantum mechanically, it would violate unitarity.
\label{f4.2}}
\end{figure}

In order to incorporate some form of feedback into the scheme, a simple
feedback mechanism such as that used in electronic circuits would
be impossible because in such a scheme, a simple feedback loop, such
as that of Fig.~\ref{f4.2} is used, and in such a loop, one has two circuit
paths feeding into one, and quantum mechanically this would violate
unitarity, because it could not be uniquely reversed. However, quantum
mechanically, there is another way to introduce feedback, and that
is through the introduction of beam splitters, which are unitary.

\section{Model of a Feedback System in Time}

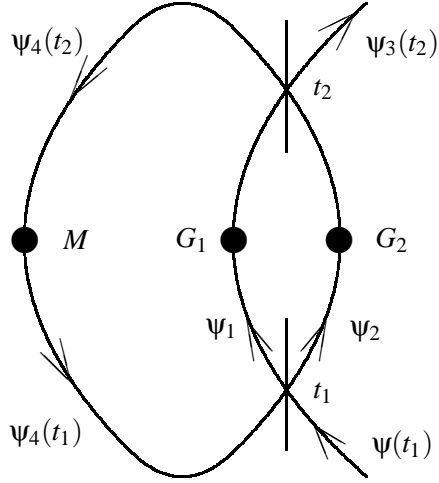
\begin{figure}
\begin{center}
\unitlength 0.70mm
\linethickness{0.7pt}
\begin{picture}(90.00,100.00)
\bezier{452}(30.00,10.00)(-10.00,50.00)(30.00,90.00)
\bezier{112}(30.00,90.00)(40.00,100.00)(50.00,90.00)
\bezier{452}(50.00,90.00)(90.00,50.00)(50.00,10.00)
\bezier{112}(50.00,10.00)(40.00,0.00)(30.00,10.00)
\bezier{544}(75.00,95.00)(24.33,50.00)(75.00,5.00)
\put(10.33,50.00){\circle*{5.20}}
\put(50.00,50.00){\circle*{5.20}}
\put(70.00,50.00){\circle*{5.20}}
\put(20.00,50.00){\makebox(0,0)[cc]{$M$}}
\put(42.00,50.00){\makebox(0,0)[cc]{$G_1$}}
\put(80.00,50.00){\makebox(0,0)[cc]{$G_2$}}
\put(60.00,91.67){\line(0,-1){25.00}}
\put(60.00,35.00){\line(0,-1){25.00}}
\put(67.00,78.67){\makebox(0,0)[cc]{$t_2$}}
\put(67.00,21.00){\makebox(0,0)[cc]{$t_1$}}
\put(82.00,11.00){\makebox(0,0)[cc]{$\psi (t_1)$}}
\put(82.00,87.33){\makebox(0,0)[cc]{$\psi_3(t_2)$}}
\put(14.67,87.67){\makebox(0,0)[cc]{$\psi_4(t_2)$}}
\put(14.33,13.33){\makebox(0,0)[cc]{$\psi_4(t_1)$}}
\put(75.00,32.67){\makebox(0,0)[cc]{$\psi_2$}}
\put(47.67,33.00){\makebox(0,0)[cc]{$\psi_1$}}
\put(52.50,34.67){\line(1,-6){1.28}}
\put(52.67,34.33){\line(5,-6){4.58}}
\put(19.00,77.00){\line(2,5){3.20}}
\put(19.00,77.00){\line(6,5){6.83}}
\put(67.50,34.67){\line(-1,-6){1.28}}
\put(67.33,34.33){\line(-5,-6){4.58}}
\put(65.33,14.50){\line(2,-5){2.47}}
\put(65.33,14.67){\line(5,-3){5.83}}
\put(19.00,23.00){\line(-1,4){2.04}}
\put(19.00,23.00){\line(-1,1){5.67}}
\put(73.05,93.19){\line(-2,-1){6.91}}
\put(73.05,93.19){\line(-1,-2){3.54}}
\end{picture}
\end{center}
    \caption{A quantum time evolution scheme with feedback. With no feedback,
$\psi\left(t_{1}\right)$ would evolve through $G_{1}$ into $\psi_{3}\left(t_{2}\right).$
There is another evolution channel $G_{2}$ and $a$ feedback channel
$M$ that alter the output at time $t_{2}$.
\label{f4.3}}
\end{figure}

The model that we introduce is one which has two beam splitters, which
allows us to generalize the classical scheme of Fig.~\ref{f4.2}, and at the
same time to present a unitary scheme allowing the particle to sample
earlier times. This should not be confused with the operation of time
reversal, which is an anti-unitary operation. The scheme is shown
in Fig.~\ref{f4.3}.

In this scheme, if there were no feedback, then the standard unitary
time development would have $\psi\left(t_{1}\right)$ evolving into
$\psi_{3}\left(t_{2}\right),$
\begin{equation}
\psi_{3}\left(t_{2}\right)=G_{1}\psi\left(t_{1}\right).\,\,\,\label{4.2}\end{equation}

Here, the operator $M$ generates the effects of the feedback in time.
These 'beam splitters' are figurative, and their role is merely to
couple the two incoming channels to two outgoing channels. The operator
$G_{1}$ represents the ordinary time development in the absence of
time feedback. The operator $G_{2}$ represents an alternate possible
time evolution that can take place and compete with $G_{1}$ because
there is feedback. We want to find $\psi_{3}\left(t_{2}\right)=f\left(\psi\left(t_{1}\right)\right)$
in the presence of the feedback in time that is generated by the operator
$M$.

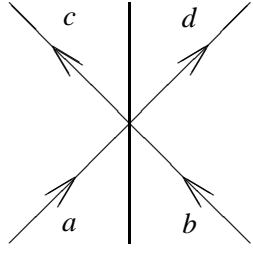
\begin{figure}
\begin{center}
\unitlength 0.80mm
\linethickness{0.7pt}
\begin{picture}(40.00,40.00)
\put(0.00,0.00){\line(1,1){40.00}}
\put(40.00,0.00){\line(-1,1){40.00}}
\put(20.00,40.00){\line(0,-1){40.00}}
\put(10.00,3.00){\makebox(0,0)[cc]{$a$}}
\put(30.00,3.33){\makebox(0,0)[cc]{$b$}}
\put(30.00,37.67){\makebox(0,0)[cc]{$d$}}
\put(10.00,37.33){\makebox(0,0)[cc]{$c$}}
\put(10.98,10.98){\line(-5,-3){4.99}}
\put(10.98,10.98){\line(-3,-5){2.93}}
\put(28.94,10.98){\line(5,-3){4.99}}
\put(28.94,10.98){\line(3,-5){2.93}}
\put(7.19,32.73){\line(5,-3){4.99}}
\put(7.19,32.73){\line(3,-5){2.93}}
\put(33.03,32.93){\line(-5,-3){4.99}}
\put(33.03,32.93){\line(-3,-5){2.93}}
\end{picture}
\end{center}
    \caption{The beam splitter transmits with an amplitude $\alpha$
and reflects with an amplitude $i\beta$. The factor of $i$ preserves
unitrarity.
\label{f4.4}}
\end{figure}

At the beam splitters, which are shown in more detail in Fig.~\ref{f4.4},
the forward amplitude is $\alpha$, while the reflected amplitude
is i$\beta$. One needs the factor of i because the two amplitudes
must differ by 90° in order to preserve unitarity. Normally, we expect
that $\alpha\gg\beta$, and in the limit $\alpha=1$, we should get
the situation represented by (\ref{4.2}).

The beam splitters perform the unitary transformation
\begin{equation}
\left|a\right\rangle =\alpha\left|d\right\rangle
+i\beta\left|c\right\rangle ,\,\,\,\,\,\,\left|b\right\rangle
=\alpha\left|c\right\rangle +i\beta\left|d\right\rangle ,\,\,\,\,\,\,
\alpha^{2}+\beta^{2}=1.\,\,\,\label{4.3}\end{equation}

Here we assume for simplicity that $\alpha$ and $\beta$ are real.
We can invert this to obtain
\begin{equation}
\left|d\right\rangle =\alpha\left|a\right\rangle -i\beta\left|b\right\rangle ,\,\,\,
\left|c\right\rangle =\alpha\left|b\right\rangle -i\beta\left|a\right\rangle .\,\,\,\label{4.4}\end{equation}

The overall governing equations can be read directly from Fig.~\ref{f4.3}.
At time $t_{2}$ the second beam splitter determines $\psi_{3}\left(t_{2}\right)$
and $\psi_{4}\left(t_{2}\right)$. We have
\begin{equation}
\psi_{3}\left(t_{2}\right)\equiv\psi_{3}^{\prime}=\left[\alpha\psi_{1}\left(t_{2}\right)-i\beta\psi_{2}\left(t_{2}\right)\right]=\alpha\psi_{1}^{\prime}-i\beta\psi_{2}^{\prime},\,\,\,\label{4.5}\end{equation}
where the prime indicates the time $t_{2}$ in the argument, and no
prime indicates the time $t_{1}$. The wave functions $\psi_{1}$and
$\psi_{2}$ are determined at time $t_{2}$ by
\begin{equation}
\psi_{1}\left(t_{2}\right)=\psi_{1}^{\prime}=G_{1}\psi_{1}\left(t_{1}\right)=G_{1}\psi_{1},\,\,\,\label{4.6}\end{equation}
\begin{equation}
\psi_{2}^{\prime}=G_{2}\psi_{2}.\,\,\,\label{4.7}\end{equation}
So that from (\ref{4.5}),
\begin{equation}
\psi_{3}^{\prime}=\alpha G_{1}\psi_{1}-i\beta G_{2}\psi_{2},\,\,\,\label{4.8}\end{equation}
and equivalently
\begin{equation}
\psi_{4}^{\prime}=\alpha G_{2}\psi_{2}-i\beta G_{1}\psi_{1}.\,\,\,\label{4.9}\end{equation}

The propagator $M$ is what produces the feedback in time, propagating
from $t_{2}$ back to $t_{1}$, so that $\psi_{4}\left(t_{1}\right)=M\psi_{4}\left(t_{2}\right),$
or
\begin{equation}
\psi_{4}=M\psi_{4}^{\prime}.\,\,\,\label{4.10}\end{equation}
At the $t_{1}$ beam splitter
\begin{eqnarray}
\psi_{1}&=&\alpha\psi-i\beta\psi_{4},\,\,\,\label{4.11}
\\
\psi_{2}&=&\alpha\psi_{4}-i\beta\psi.\,\,\,\label{4.12}\end{eqnarray}

\section{The Solution}

First, we want to eliminate the $\psi_{4}$ ub (\ref{4.11}) and (\ref{4.12}).,
to get equations for $\psi_{1}$and $\psi_{2}$. Then from (\ref{4.8}) we
can obtain $\psi_{3}^{\prime}$. From (\ref{4.9}) and (\ref{4.10}),\begin{equation}
\psi_{4}=M\psi_{4}^{\prime}=\alpha MG_{2}\psi_{2}-i\beta MG_{1}\psi_{1}.\,\,\,\label{4.13}\end{equation}

We plug this into (\ref{4.11}) and (\ref{4.12}),
\begin{eqnarray}
\psi_{1}&=&\alpha\psi-i\beta\left(\alpha MG_{2}\psi_{2}-i\beta MG_{1}\psi_{1}\right),\,\,\,\,\label{4.14}\\
\psi_{2}&=&\alpha\left(\alpha MG_{2}\psi_{2}-i\beta MG_{1}\psi_{1}\right)-i\beta\psi.\,\,\,\label{4.15}\end{eqnarray}

We can rewrite these as
\begin{eqnarray}
\psi_{1}&=&\left(1+\beta^{2}MG_{1}\right)^{-1}\left(-i\alpha\beta MG_{2}\right)\psi_{2}+\alpha\left(1+\beta^{2}MG_{1}\right)^{-1}\psi,\,\,\,\label{4.16}
\\
\psi_{2}&=&\left(1-\alpha^{2}MG_{2}\right)^{-1}\left(-i\alpha\beta MG_{1}\right)\psi_{1}-i\beta\left(1-\alpha^{2}MG_{2}\right)^{-1}\psi .\,\,\,\,\label{4.17}\end{eqnarray}

These are two simultaneous equations that we must solve to find $\psi_{1}$
and $\psi_{2}$ as functions of $\psi$. To solve for $\psi_{1},$
substitute (\ref{4.17}) into (\ref{4.16}).
\begin{eqnarray}
\psi_{1} & = & \left(1+\beta^{2}MG_{1}\right)^{-1}\left(-i\alpha\beta MG_{2}\right)\left[\left(1-\alpha^{2}MG_{2}\right)^{-1}\left(-\alpha\beta MG_{1}\right)\psi_{1}\right.\nonumber \\
 &  & -i\beta\left.\left(1-\alpha^{2}MG\right)^{-1}\psi\right]+\alpha\left(1+\beta^{2}MG_{1}\right)^{-1}\psi.\,\,\,\label{4.18}\end{eqnarray}

This can be rewritten as
\begin{eqnarray}
\left[1+\alpha^{2}\beta^{2}\left(1+\beta^{2}MG_{1}\right)^{-1}\left(MG_{2}\right)\left(1-\alpha^{2}MG_{2}\right)^{-1}\left(MG_{1}\right)\right]\psi_{1}\nonumber \\
=\left(1+\beta^{2}MG_{1}\right)^{-1}\left[-\alpha\beta^{2}MG_{2}\left(1-\alpha^{2}MG_{2}\right)^{-1}+\alpha\right]\psi.\,\,\,\label{4.19}\end{eqnarray}

If we write this as
\begin{equation}
\left[X\right]\psi_{1}=Y^{-1}\left[Z\right]\psi,\,\,\,\label{4.20}\end{equation}
then we can simplify the equation as follows:
\begin{eqnarray}
YX & = & 1+\beta^{2}MG_{1}+\alpha^{2}\beta^{2}MG_{2}\left(1-\alpha^{2}MG_{2}\right)^{-1}MG_{1}\nonumber \\
 & = & 1+\beta^{2}\left[1+\left(1-\alpha^{2}MG_{2}\right)^{-1}\alpha^{2}MG_{2}\right]MG_{1}\nonumber \\
 & = & 1+\beta^{2}\left(1-\alpha^{2}MG_{2}\right)^{-1}MG_{1}.\,\,\,\label{4.21}\end{eqnarray}
and
\begin{eqnarray}
Z & = & \alpha\left(1-\alpha^{2}MG_{2}\right)^{-1}\left(1-\alpha^{2}MG_{2}-\beta^{2}MG_{2}\right)\nonumber \\
 & = & \alpha\left(1-\alpha^{2}MG_{2}\right)^{-1}\left(1-MG_{2}\right).\,\,\,\label{4.22}\end{eqnarray}

Thus,
\begin{equation}
\psi_{1}=\alpha\left[1+\beta^{2}\left(1-\alpha^{2}MG_{2}\right)^{-1}MG_{1}\right]^{-1}\left(1-\alpha^{2}MG_{2}\right)^{-1}\left(1-MG_{2}\right)\psi.\label{4.23}\end{equation}

Then, using the identity $A^{-1}B^{-1}=\left(BA\right)^{-1}$, we
finally get
\begin{equation}
\psi_{1}=\alpha\left(1-\alpha^{2}MG_{2}+\beta^{2}MG_{1}\right)^{-1}\left(1-MG_{2}\right)\psi.\,\,\,\label{4.24}\end{equation}

We can solve for $\psi_{2}$ similarly, by substituting (\ref{4.16}) into
(\ref{4.17}),
\begin{equation}
\psi_{2}=-i\beta\left(1-\alpha^{2}MG_{2}+\beta^{2}MG_{1}\right)^{-1}\left(1+MG_{1}\right)\psi.\,\,\,\label{4.25}\end{equation}

Notice that in the denominator term in both (\ref{4.24}) and (\ref{4.25}), $\alpha$
and $\beta$ have reversed the role of the operators they apply to.
we can finally use (\ref{4.8}) to solve for $\psi_{3}^{\prime}=\psi_{3}\left(t_{2}\right),$
\begin{equation}
\psi_{3}\left(t_{2}\right)=\left[\alpha^{2}G_{1}D\left(1-MG_{2}\right)-\beta^{2}G_{2}D\left(1+MG_{1}\right)\right]\psi\left(t_{1}\right),\,\,\,\label{4.26}\end{equation}
where $D=\left(1+\beta^{2}MG_{1}-\alpha^{2}MG_{2}\right)^{-1}.$

\section{Some Important Special Cases}

{\bf The Case} $\alpha=1$, $\beta=0.$ This is the case where there is no
feedback. Here
\begin{equation}
\psi_{3}^{\prime}=G_{1}\left(1-MG\right)^{-1}\left(1-MG_{2}\right)\psi=G_{1}\psi\,\,\,\label{4.27}\end{equation}

{\bf The Case} $\beta=1$, $\alpha=0.$ This is the case where there is only
feedback. Here
\begin{equation}
\psi_{3}^{\prime}=-G_{2}\left(1+MG_{1}\right)^{-1}\left(1+MG_{1}\right)\psi=-G_{2}\psi.\,\,\,\label{4.28}\end{equation}

{\bf The Case} $G_{2}=-G_{1}$.
This corresponds to the case where both paths lead to the same future.
(the ``$-$'' sign is due to the phasing effect of the beam splitters.)

\begin{eqnarray}
\psi_{3}^{\prime}
&=&
\left[
\alpha^2 G_1 D(1+MG_1)
+\beta^2 G_1 D(1+MG_1)
\right] \psi ,\nonumber   \\
D&=& \left(
1+\beta^2 MG_1+\alpha^2 MG_1
\right)^{-1}
=
\left(
1+MG_1
\right)^{-1}
.
\label{4.29}\end{eqnarray}
Then
\begin{equation}
\psi_{3}^{\prime}=G_1\psi ,\,\,\,\label{4.30}\end{equation}
as we would expect.

{\bf The Case} $\beta \ll 1$. This is expected to be the usual case. Then
the answer only depends on $\beta^{2}=\gamma.$ Also, $\alpha^{2}=1-\beta^{2}=1-\gamma$.
Then to lowest order in $\gamma$, the denominator $D$ in (\ref{4.26})
becomes
\begin{eqnarray}
D & = & \left[1+\gamma MG_{1}-\left(1-\gamma\right)MG_{2}\right]^{-1}\nonumber \\
 & = & \left(1-MG_{2}\right)^{-1}-\gamma\left(1-MG_{2}\right)^{-1}\left(MG_{1}+MG_{2}\right)\left(1-MG_{2}\right)^{-1},\label{4.31}\end{eqnarray}

so that
\begin{eqnarray}
\psi_{3}^{\prime} & = & \left\{ \left(1-\gamma\right)G_{1}\left[1-\gamma\left(1-MG_{2}\right)^{-1}\left(MG_{1}+MG_{2}\right)\right]\right.\nonumber \\
 &  & \left.-\gamma G_{2}\left(1-MG_{2}\right)^{-1}\left(1+MG_{1}\right)\right\} \psi\nonumber \\
 & = & G_{1}\psi-\gamma\left(G_{1}+G_{2}\right)\left(1-MG_{2}\right)\left(1+MG_{1}\right)\psi.\,\,\,\label{4.32}\end{eqnarray}

\section{The Classical Paradox of Shooting your Father}

The most interesting case is the one that corresponds to the classical
paradox where you shoot your father before he has met your mother,
so that you can never be born. This case has a rather fascinating
quantum-mechanical resolution.
Actually, there are two possible realizations of this case.
The first is the case $G_{1}=0$, where
there is a perfect absorber in the beam so that the system without
any feedback would never get to evolve to time $t_{2}.$ But quantum
mechanically, we assume that there is another path along $G_{2}$,
the one where you do not shoot your father, that has a probability
$\beta$ without feedback. In quantum theory we deal with probabilities,
and as long as there is any chance that you may not meet your father,
we must take this into account.

The solution in this case is
\begin{equation}
\psi_{3}^{\prime}=-\beta^{2}G_{2}\left(1-\alpha^{2}MG_{2}\right)^{-1}\psi.\,\,\,\label{4.33}\end{equation}

We assume for simplicity that $G_{2}$ is just the standard time evolution
operator
\begin{equation}
G_{2}=e^{-iE\left(t_{2}-t_{1}\right)/h},\,\,\,\label{4.34}\end{equation}
and $M$ is just the simplest backwards in time evolution operator
\begin{equation}
M=e^{-iE\left(t_{1}-t_{2}\right)/h+i\phi},\,\,\,\label{4.35}\end{equation}
where we have also allowed for an extra phase shift. Then
\begin{equation}
\psi_{3}^{\prime}=-\beta^{2}e^{-iE\left(t_{2}-t_{1}\right)/h}\left(1-\alpha^{2}e^{i\phi}\right)^{-1}\psi,\,\,\,\label{4.36}\end{equation}

\begin{equation}
\left|\psi_{3}^{\prime}\right|^{2}=
{\beta^{4}\over \left(1-\alpha^{2}e^{i\phi}\right)\left(1-\alpha^{2}e^{-i\phi}\right)}
\left|\psi\right|^{2}
={1 \over 1+4\left(\alpha^{2}/\beta^{4}\right)\sin^{2}\left(\phi/2\right)}
\left|\psi\right|^{2}.\,\,\,\label{4.37}
\end{equation}

Note that for $\phi=0,\,\psi_{3}^{\prime}=-e^{-iE\Delta t/h}\psi,$
for any value of $\beta$. That means that no matter how small the
probability of your ever having reached here in the first place, the
fact that you are here, which can only happen because $\alpha\neq1$,
guarantees that even though you are certain to have shot your father
if you had met him $\left(G_{1}=0\right),$ nonetheless you will not
have met him! You will have taken the other path, with 100\% certainly.
Obviously, this must be the case, if you are to be here at all.

How can we understand this result? In our model, with $\phi=0$, we
have $G_{1}=0,$ and $MG_{2}=1$. Also, we will assume that $\beta\ll1$,
even though this is not necessary. The various amplitudes are
\begin{equation}
\left|\psi_{1}\right|=0,\,\,\,\left|\psi_{2}/\psi\right|=1/\beta,\,\,\,\left|\psi_{4}/\psi\right|=\alpha/\beta,\,\,\,\left|\psi_{3}^{\prime}/\psi\right|=1.\,\,\,\label{4.38}\end{equation}

So we see that the two paths of the beam splitter at $t_{1}$ leading
to the path $\psi_{1}$ cancel out. But of the original beam $\psi,$
$\alpha$ passes through to $\psi_{1}$, while of the beam $\psi_{4}$,
only the fraction $\beta$ leaks through to $\psi_{1}$. So the beam
$\psi_{4}$ must have a very large amplitude, which it does, as we
can see from (\ref{4.38}), so that the two contributions can cancel at $\psi_{1}.$
In fact $\psi_{4}$ has a much larger amplitude than the original
beam! Similarly, in order to have $\left|\psi_{3}^{\prime}\right|=\left|\psi\right|$,
then
$\psi_{2}$ must have a very large amplitude. Thus we see that there
is a large current flowing around the system, between $\psi_{2}$
and $\psi_{4}$. But does this not violate unitarity? The answer is
that if they were both running forward in time, it would. But one
of these currents is running forward in time, while the other runs
backward in time, and so they do not in this case violate unitarity.
This is how our solution is possible.

There is a second possible way to bring about this case, namely to allow any $G_1$, but to make
$M=-G_1^{-1}$,
so that the purpose of the time travel will be to undo $G_1$.
(Again the ``$-$'' sign is due to the phasing of the beam splitters.)
Then
\begin{eqnarray}
\psi_3^\prime
&=&
\left[
\alpha^2 G_1D(1+G_1^{-1}G_2)+0
\right]\psi ,
\nonumber
\\
D
&=&
\left(
1-\beta^2+\alpha^2 G_1^{-1}G_2
\right)^{-1}=
\left(
{1\over \alpha^2}
\right)
\left(
1+G_1^{-1}G_2
\right)^{-1}.
\end{eqnarray}
Thus,
\begin{equation}
\psi_3^\prime =G_1\psi ,
\end{equation}
and instead of undoing $G_1$, $M$ wipes out the alternative possible future,
thus guaranteeing the future that has already happened.

\section{Conclusion}

According to our model, if you travel into the past quantum mechanically,
you would only see those alternatives consistent with the world you
left behind you. In other words, while you are aware of the past,
you cannot change it. No matter how unlikely the events are that could
have led to your present circumstances, once they have actually occurred,
they cannot be changed. Your trip would set up resonances that are
consistent with the future that has already unfolded.

This also has enormous consequences on the paradoxes of free will.
It shows that it is perfectly logical to assume that one has many
choices and that one is free to take any one of them. Until a choice
is taken, the future is not determined. However, once a choice is
taken, and it leads to a particular future, it was inevitable. It
could not have been otherwise. The boundary conditions that the future
events happen as they already have, guarantees that they must have
been prepared for in the past. So, looking backwards, the world is
deterministic. However, looking forwards, the future is probabilistic.
This completely explains the classical paradox. In fact, it serves
as a kind of indirect evidence that such feedback must actually take
place in nature, in the sense that without it, a paradox exists, while
with it, the paradox is resolved. (Of course, there is an equally
likely explanation, namely that going backward in time is impossible.
This also solves the paradox by avoiding it.)

The model also has consequences concerning the many-worlds interpretation
of quantum theory. The world may appear to keep splitting so far as
the future is concerned. However, once a measurement is made, only
those histories consistent with that measurement are possible. In
other words, with time travel, other alternative worlds do not exist,
as once a measurement has been made confirming the world we live in,
the other worlds would be impossible to reach from the original one.
This explanation makes the von Neumann state reduction hypothesis
much more reasonable, and in fact acts as a sort of justification
of it.

Another interesting point comes from examining (\ref{4.37}). For small angles
$\phi$, we see that
\begin{equation}
\left|\psi_{3}^{\prime}\right|^{2}=\frac{1}{1+4\left(\alpha^{2}/\beta^{4}\right)\sin^{2}\left(\phi /2\right)}
\left|\psi\right|^{2}\rightarrow\frac{1}{1+\alpha^{2}\phi^{2}/\beta^{4}}\left|\psi\right|^{2},\,\,\,\label{4.39}
\end{equation}
so that the above result is strongly resonant, with a Lorentzian shape,
and a width $\Delta\phi\sim\beta^{2}$, since $\alpha\sim1.$Thus
less 'deterministic' and fuzzier time traveling might be possible,
a possibility we have not yet explored. Neither have we explored the
possibility that feedback should be possible into the future as well
as the past. Of course in this case, it ought to be called 'feedforward'
- rather than feedback.

\section*{References}

[1] There are many books about the nature of time. The two main paradoxes
are the question of reversing the direction of time so that time travel
is possible, and the issue of the 'arrow of time', namely, why time
flows in one direction. A book of interesting essays on both questions
is \textit{The Nature of Time}, ed. by R. Flood and M. Lockwood, Basil
Blackwell, Cambridge, Mass. (1986). Some other interesting references
are: \textit{Time's Arrow and Archimedes' Point}, by H. Price, Oxford
University Press, New York (1996); \textit{The Physical Basis of the
Direction of Time}, by H.D. Zeh, Springer Verlag, Berlin (1999); and
\textit{Time's Arrows and Quantum Measurement}, by L.S. Schulmann,
Cambridge University Press, Cambridge (1997).

[2] This paper is an expanded version of an earlier paper on the subject,
D.M. Greenberger and K. Svozil, in: \textit{Between Chance and Choice}, ed.
by H. Atmanspacher and R. Bishop, Imprint Academic, Thorverton England
(2002), pp. 293-308.

[3] This paper contains minor changes to our paper published as Chapter 4 of
\textit{Quo Vadis Quantum Mechanics?}, ed. by
A. Elitzur, S. Dolev and N. Kolenda, Springer Verlag, Berlin (2005), pp. 63-72.

[4] We have just become aware of an article by
D. T. Pegg in \textit{Times' arrows, quantum measurement and superluminal behavior},
ed. by
 D. Mugnai, A. Ranfagni and L. S. Schulman,
Consiglio Nazionale Delle Richerche, Roma, (2001) p. 113, eprint quant-ph/0506141,
in which related ideas have been developed.
Self-consistent causal loops arising from backward time travel have been discussed
by K. Svozil, in \textit{Fundamental Problems in Quantum Theory: A Conference Held in
Honor of Professor John A. Wheeler}, ed. by D. M. Greenberger and A.
Zeilinger, \textit{Annals of the New York Academy of Sciences 755} (1995), pp. 834-841,
eprint quant-ph/9502008.

\end{document}